\pdfoutput=1

\documentclass[12pt,a4paper]{article}

\usepackage{ifthen} 
\newboolean{pdflatex}
\setboolean{pdflatex}{true} 

\newboolean{articletitles}
\setboolean{articletitles}{true} 

\newboolean{uprightparticles}
\setboolean{uprightparticles}{false} 

\newboolean{inbibliography}
\setboolean{inbibliography}{false} 

\usepackage{microtype}
\usepackage{lineno}  
\usepackage{xspace} 

\usepackage{graphicx}  
\usepackage{color}
\usepackage{colortbl}
\graphicspath{{./figs/}} 

\usepackage{amsmath} 
\usepackage{amssymb}
\usepackage{amsfonts}
\usepackage{upgreek} 

\newcommand*\patchAmsMathEnvironmentForLineno[1]{%
\expandafter\let\csname old#1\expandafter\endcsname\csname #1\endcsname
\expandafter\let\csname oldend#1\expandafter\endcsname\csname
end#1\endcsname
 \renewenvironment{#1}%
   {\linenomath\csname old#1\endcsname}%
   {\csname oldend#1\endcsname\endlinenomath}%
}
\newcommand*\patchBothAmsMathEnvironmentsForLineno[1]{%
  \patchAmsMathEnvironmentForLineno{#1}%
  \patchAmsMathEnvironmentForLineno{#1*}%
}
\AtBeginDocument{%
\patchBothAmsMathEnvironmentsForLineno{equation}%
\patchBothAmsMathEnvironmentsForLineno{align}%
\patchBothAmsMathEnvironmentsForLineno{flalign}%
\patchBothAmsMathEnvironmentsForLineno{alignat}%
\patchBothAmsMathEnvironmentsForLineno{gather}%
\patchBothAmsMathEnvironmentsForLineno{multline}%
}

\usepackage{hyperref}    
\usepackage[all]{hypcap} 

\def\lhcb {LHCb\xspace}
\def\ux85 {UX85\xspace}

\ifthenelse{\boolean{uprightparticles}}%
{

 \def\Ppi         {\ensuremath{\uppi}\xspace}

 \def\Ppsi        {\ensuremath{\uppsi}\xspace}

 \def\PDelta      {\ensuremath{\Delta}\xspace}                 
 \def\PXi      {\ensuremath{\Xi}\xspace}                 
 \def\PLambda      {\ensuremath{\Lambda}\xspace}                 
 \def\PSigma      {\ensuremath{\Sigma}\xspace}                 
 \def\POmega      {\ensuremath{\Omega}\xspace}                 
 \def\PUpsilon      {\ensuremath{\Upsilon}\xspace}                 
 
 \def\PB      {\ensuremath{\mathrm{B}}\xspace}                 
                  
 \def\PD      {\ensuremath{\mathrm{D}}\xspace}

 \def\PJ      {\ensuremath{\mathrm{J}}\xspace}                 
 \def\PK      {\ensuremath{\mathrm{K}}\xspace}

 \def\Pb      {\ensuremath{\mathrm{b}}\xspace}                 
 \def\Pc      {\ensuremath{\mathrm{c}}\xspace}

 \def\Pi      {\ensuremath{\mathrm{i}}\xspace}

 \def\Ps      {\ensuremath{\mathrm{s}}\xspace}

}
{

 \def\Ppi         {\ensuremath{\pi}\xspace}

 \def\Ppsi        {\ensuremath{\psi}\xspace}                 
                  
 \mathchardef\PDelta="7101
 \mathchardef\PXi="7104
 \mathchardef\PLambda="7103
 \mathchardef\PSigma="7106
 \mathchardef\POmega="710A
 \mathchardef\PUpsilon="7107
                  
 \def\PB      {\ensuremath{B}\xspace}                 
                  
 \def\PD      {\ensuremath{D}\xspace}

 \def\PJ      {\ensuremath{J}\xspace}                 
 \def\PK      {\ensuremath{K}\xspace}

 \def\Pb      {\ensuremath{b}\xspace}                 
 \def\Pc      {\ensuremath{c}\xspace}

 \def\Pi      {\ensuremath{i}\xspace}

 \def\Ps      {\ensuremath{s}\xspace}

}

\def\squark    {\ensuremath{\Ps}\xspace}

\def\cquark    {\ensuremath{\Pc}\xspace}

\def\bquark    {\ensuremath{\Pb}\xspace}

\def\pion  {\ensuremath{\Ppi}\xspace}

\def\pip   {\ensuremath{\pion^+}\xspace}
\def\pim   {\ensuremath{\pion^-}\xspace}

\def\kaon  {\ensuremath{\PK}\xspace}
  \def\Kbar  {\kern 0.2em\overline{\kern -0.2em \PK}{}\xspace}

\def\Kz    {\ensuremath{\kaon^0}\xspace}
\def\Kzb   {\ensuremath{\Kbar^0}\xspace}
\def\KzKzb {\ensuremath{\Kz \kern -0.16em \Kzb}\xspace}
\def\Kp    {\ensuremath{\kaon^+}\xspace}
\def\Km    {\ensuremath{\kaon^-}\xspace}

\def\KpKm  {\ensuremath{\Kp \kern -0.16em \Km}\xspace}

  \def\Dbar    {\kern 0.2em\overline{\kern -0.2em \PD}{}\xspace}
\def\D       {\ensuremath{\PD}\xspace}

\def\Dz      {\ensuremath{\D^0}\xspace}
\def\Dzb     {\ensuremath{\Dbar^0}\xspace}
\def\DzDzb   {\ensuremath{\Dz {\kern -0.16em \Dzb}}\xspace}
\def\Dp      {\ensuremath{\D^+}\xspace}
\def\Dm      {\ensuremath{\D^-}\xspace}

\def\DpDm    {\ensuremath{\Dp {\kern -0.16em \Dm}}\xspace}

\def\B       {\ensuremath{\PB}\xspace}
  \def\Bbar    {\kern 0.18em\overline{\kern -0.18em \PB}{}\xspace}
\def\Bb      {\ensuremath{\Bbar}\xspace}

\def\Bzb     {\ensuremath{\Bbar^0}\xspace}

\def\Bsb     {\ensuremath{\Bbar^0_\squark}\xspace}
\def\Bdb     {\ensuremath{\Bbar^0}\xspace}

\def\jpsi     {\ensuremath{{\PJ\mskip -3mu/\mskip -2mu\Ppsi\mskip 2mu}}\xspace}

  \def\Y#1S{\ensuremath{\PUpsilon{(#1S)}}\xspace}

\def\Lbar {\ensuremath{\kern 0.1em\overline{\kern -0.1em\Lambda\kern -0.05em}\kern 0.05em{}}\xspace}

\def\to                 {\ensuremath{\rightarrow}\xspace}

\def\AT#1     {\ensuremath{A_{\mathrm{T}}^{#1}}\xspace}           

\def\C#1      {\ensuremath{\mathcal{C}_{#1}}\xspace}                       
\def\Cp#1     {\ensuremath{\mathcal{C}_{#1}^{'}}\xspace}                    
\def\Ceff#1   {\ensuremath{\mathcal{C}_{#1}^{\mathrm{(eff)}}}\xspace}        
\def\Cpeff#1  {\ensuremath{\mathcal{C}_{#1}^{'\mathrm{(eff)}}}\xspace}       
\def\Ope#1    {\ensuremath{\mathcal{O}_{#1}}\xspace}                       
\def\Opep#1   {\ensuremath{\mathcal{O}_{#1}^{'}}\xspace}                    

\newcommand{\ket}[1]{\ensuremath{|#1\rangle}}              

\newcommand{\tev}{\ensuremath{\mathrm{\,Te\kern -0.1em V}}\xspace}
\newcommand{\gev}{\ensuremath{\mathrm{\,Ge\kern -0.1em V}}\xspace}
\newcommand{\mev}{\ensuremath{\mathrm{\,Me\kern -0.1em V}}\xspace}
\newcommand{\kev}{\ensuremath{\mathrm{\,ke\kern -0.1em V}}\xspace}
\newcommand{\ev}{\ensuremath{\mathrm{\,e\kern -0.1em V}}\xspace}
\newcommand{\gevc}{\ensuremath{{\mathrm{\,Ge\kern -0.1em V\!/}c}}\xspace}
\newcommand{\mevc}{\ensuremath{{\mathrm{\,Me\kern -0.1em V\!/}c}}\xspace}
\newcommand{\gevcc}{\ensuremath{{\mathrm{\,Ge\kern -0.1em V\!/}c^2}}\xspace}
\newcommand{\gevgevcccc}{\ensuremath{{\mathrm{\,Ge\kern -0.1em V^2\!/}c^4}}\xspace}
\newcommand{\mevcc}{\ensuremath{{\mathrm{\,Me\kern -0.1em V\!/}c^2}}\xspace}

\def\mum  {\ensuremath{\,\upmu\rm m}\xspace}

\newcommand{\chisq}{\ensuremath{\chi^2}\xspace}

\def\gsim{{~\raise.15em\hbox{$>$}\kern-.85em
          \lower.35em\hbox{$\sim$}~}\xspace}
\def\lsim{{~\raise.15em\hbox{$<$}\kern-.85em
          \lower.35em\hbox{$\sim$}~}\xspace}

\def\pt         {\mbox{$p_{\rm T}$}\xspace}

\def\evtgen     {\mbox{\textsc{EvtGen}}\xspace}
\def\pythia     {\mbox{\textsc{Pythia}}\xspace}

\def\geant      {\mbox{\textsc{Geant4}}\xspace}

\def\tell1  {TELL1\xspace}
\def\ukl1   {UKL1\xspace}

\def\f1 {\ensuremath{f_1(1285)}\xspace}
\usepackage{cite} 
\usepackage{setspace}
\usepackage{mciteplus}

\begin{document}

\renewcommand{\thefootnote}{\fnsymbol{footnote}}
\setcounter{footnote}{1}

\begin{titlepage}
\pagenumbering{roman}

\vspace*{-1.5cm}
\centerline{\large EUROPEAN ORGANIZATION FOR NUCLEAR RESEARCH (CERN)}
\vspace*{1.5cm}
\hspace*{-5mm}\begin{tabular*}{16cm}{lc@{\extracolsep{\fill}}r}
\vspace*{-12mm}\mbox{\!\!\!\includegraphics[width=.12\textwidth]{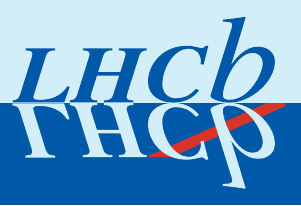}}& & \\ 
 & & CERN-PH-EP-2013-186\\
 & & LHCb-PAPER-2013-055\\  
 & &8 October 2013 \\ 
 & & \\
\end{tabular*}

\vspace*{3.0cm}

{\bf\boldmath\huge
\begin{center}
Observation of $\Bdb_{(s)}\to\jpsi f_1(1285)$ decays and measurement of the $f_1(1285)$  mixing angle
\end{center}
}

\vspace*{2.0cm}

\begin{center}
The LHCb collaboration\footnote{Authors are listed on the following pages.}
\end{center}

\vspace{\fill}

\begin{abstract}
  \noindent
Decays of \Bsb and \Bdb mesons  into $\jpsi  \pi^+\pi^-\pi^+\pi^-$ final states, produced in $pp$ collisions at the LHC, are investigated using data corresponding to an integrated luminosity of  3~fb$^{-1}$ collected with the LHCb detector. $\Bzb_{(s)}\to\jpsi f_1(1285)$ decays  are seen for the
first time, and the branching fractions are measured. Using these
rates, the \f1 mixing angle between strange and non-strange components of
its wave function in the $q\overline{q}$ structure model is determined to be $\pm(24.0^{\,+3.1\,+0.6}_{\,-2.6\,-0.8})^{\circ}$.
Implications on the possible tetraquark nature of the \f1 are discussed.
\end{abstract}

\vspace*{2.0cm}

\begin{center}
  Submitted to Phys.~Rev.~Lett. 
\end{center}

\vspace{\fill}

{\footnotesize 
\centerline{\copyright~CERN on behalf of the \lhcb collaboration, license \href{http://creativecommons.org/licenses/by/3.0/}{CC-BY-3.0}.}}
\vspace*{2mm}

\end{titlepage}

\newpage
\setcounter{page}{2}
\mbox{~}

\centerline{\large\bf LHCb collaboration}
\begin{flushleft}
\small
R.~Aaij$^{40}$, 
B.~Adeva$^{36}$, 
M.~Adinolfi$^{45}$, 
C.~Adrover$^{6}$, 
A.~Affolder$^{51}$, 
Z.~Ajaltouni$^{5}$, 
J.~Albrecht$^{9}$, 
F.~Alessio$^{37}$, 
M.~Alexander$^{50}$, 
S.~Ali$^{40}$
% ...

{\footnotesize \it
$ ^{1}$Centro Brasileiro de Pesquisas F\'{i}sicas (CBPF), Rio de Janeiro, Brazil\\
% ...
$ ^{5}$Clermont Universit\'{e}, Universit\'{e} Blaise Pascal, CNRS/IN2P3, LPC, Clermont-Ferrand, France\\
$ ^{6}$CPPM, Aix-Marseille Universit\'{e}, CNRS/IN2P3, Marseille, France\\
% ...
$ ^{9}$Fakult\"{a}t Physik, Technische Universit\"{a}t Dortmund, Dortmund, Germany\\
% ...
$ ^{36}$Universidad de Santiago de Compostela, Santiago de Compostela, Spain\\
$ ^{37}$European Organization for Nuclear Research (CERN), Geneva, Switzerland\\
% ...
$ ^{40}$Nikhef National Institute for Subatomic Physics, Amsterdam, The Netherlands\\
% ...
$ ^{45}$H.H. Wills Physics Laboratory, University of Bristol, Bristol, United Kingdom\\
% ...
$ ^{50}$School of Physics and Astronomy, University of Glasgow, Glasgow, United Kingdom\\
$ ^{51}$Oliver Lodge Laboratory, University of Liverpool, Liverpool, United Kingdom\\
% ...
$ ^{q}$Universit\`{a} di Padova, Padova, Italy\\
% ...
}
\end{flushleft}


\cleardoublepage

\renewcommand{\thefootnote}{\arabic{footnote}}
\setcounter{footnote}{0}



\pagestyle{plain} 
\setcounter{page}{1}
\pagenumbering{arabic}  
\noindent 
 Light flavorless hadrons, $f$, are not entirely understood as $q\overline{q}$ states. Some states with the
same quantum numbers such as the $\eta$ and $\eta'$ exhibit mixing \cite{PDG}.  Others, such
as the $f_0(500)$ and the $f_0(980)$, could be mixed $q\overline{q}$ states, or they could be comprised of
tetraquarks \cite{Fariborz:2009cq,*Weinberg:2013cfa,*Hooft:2008we,*Achasov:2012kk}. In addition some states, such as the $f_0(1500)$, are discussed as being made
solely of gluons \cite{Ochs:2013gi,*Jaffe:1976ig}. Understanding if the $f$ states are indeed explained by the quark model
is crucial to identifying other exotic structures. Previous investigations of \Bsb and \Bdb 
decays (called generically $\Bbar$) into a \jpsi meson and a $\pi^+\pi^-$ \cite{LHCb:2012ae,Aaij:2013zpt} or $K^+K^-$ \cite{Aaij:2013orb,Aaij:2013mtm} pair have revealed the presence
of several light flavorless meson resonances including the $f_0(500)$ and the $f_0(980)$. Use of
$\Bbar\to \jpsi f$ decays has been suggested as an excellent way of both measuring mixing
angles and discerning if some of the $f$ states are  tetraquarks \cite{Stone:2013eaa,Fleischer:2011au,*Fleischer:2011ib}. In this Letter the $\jpsi  \pi^+\pi^-\pi^+\pi^-$ final state is investigated with the aim of seeking additional $f$ states. (Mention of a particular process also implies the use of its charge conjugated decay.)

Data are obtained from  3~fb$^{-1}$ of integrated luminosity collected with the LHCb
detector \cite{LHCb-det} using $pp$ collisions. One third of the data was acquired at a center-of-mass
energy of 7 TeV, and the remainder at 8 TeV. The \lhcb detector is a single-arm forward
spectrometer covering the \mbox{pseudorapidity} range $2<\eta <5$, designed
for the study of particles containing \bquark or \cquark quarks. The detector includes a high precision tracking system consisting of a
silicon-strip vertex detector surrounding the $pp$ interaction region,
a large-area silicon-strip detector located upstream of a dipole
magnet with a bending power of about $4{\rm\,Tm}$, and three stations
of silicon-strip detectors and straw drift tubes placed
downstream. The combined tracking system provides a momentum measurement with
relative uncertainty that varies from 0.4\% at 5~GeV to 0.6\% at 100~GeV. (We work in units where $c$=1.)
The impact parameter (IP) is defined as the minimum track distance with respect to the primary vertex. For tracks with large transverse momentum, \pt, with respect to the proton beam direction, the IP  resolution is approximately 20\mum. Charged hadrons are identified using two ring-imaging Cherenkov (RICH) detectors. Photon, electron and hadron candidates are identified by a calorimeter system consisting of scintillating-pad and pre-shower detectors, an electromagnetic calorimeter and a hadronic calorimeter. Muons are identified by a system composed of alternating layers of iron and multiwire proportional chambers. 
 
 The LHCb trigger \cite{Aaij:2012me} consists of a hardware stage, based on information from the
calorimeter and muon systems, followed by a software stage that applies event
reconstruction. Events selected for this analysis are triggered by a  candidate $\jpsi\to\mu^+\mu^-$ decay,
required to be consistent with coming from the decay of a $b$-hadron by using either IP requirements or detachment from the
associated primary vertex. Simulations are performed using \pythia \cite{Sjostrand:2006za} with the specific
tuning given in Ref. \cite{LHCb-PROC-2011-005}, and the LHCb detector description based on \geant \cite{Allison:2006ve,*Agostinelli:2002hh}
described in Ref. \cite{LHCb-PROC-2011-006}. Decays of $b$-hadrons are based on \evtgen \cite{Lange:2001uf}.
 
 Events are preselected and then are further filtered using a multivariate analyzer based
on the boosted decision tree (BDT) technique \cite{Breiman}. In the preselection, all charged track
candidates are required to have \pt $>$ 250 MeV, while for muon candidates the
requirement is \pt $>$ 550 MeV. Events must have a $\mu^+\mu^-$ combination that forms
a common vertex with $\chi^2 < 20$, an invariant mass between $-48$ and +43 MeV of
 the \jpsi meson mass, and are constrained to the \jpsi mass. The four pions must have a vector summed $\pt >1$~GeV,  form a vertex with $\chi^2 <50$ for five degrees of freedom,
and a common vertex with the \jpsi candidate with $\chi^2 < 90$ for nine degrees of freedom.
The angle between the \Bb  momentum and the vector from the primary vertex to the \Bb decay vertex is required to be  smaller than 2.56$^{\circ}$. Particle identification \cite{Adinolfi:2012an}
requirements are based on the difference in the logarithm of the likelihood, DLL$(h_1-h_2)$,
to distinguish between the hypotheses $h_1$ and $h_2$. We require DLL$(\pi-\mu) > -10 $ and
DLL$(\pi-K) > -10$. We also explicitly eliminate candidate $\psi (2S)[$or $X(3872)]\to\jpsi\pi^+\pi^-$ 
events by rejecting any candidate where one $\jpsi\pi^+\pi^-$ combination is within 23 MeV of
the $\psi (2S)$  or 9 MeV of the $X(3872)$ meson masses. Other resonant contributions such as
 $\Bbar\to \psi(4160)\pi^+\pi^-$ are searched for, but not found.
 
 The BDT uses 12 variables that are chosen to separate signal and background: the
minimum DLL$(\pi-\mu)$ of the $\mu^+$ and $\mu^-$, the scalar \pt sum of the four pions, and the vector
\pt sum of the four pions; relating to the $\Bbar$ candidate: the flight distance, the vertex $\chi^2$, the \pt, and
the $\chi^2_{\rm IP}$, which is defined as the difference in \chisq of a given primary vertex reconstructed with and without the considered particle. In addition, considering the $\pi^+\pi^+$ and $\pi^-\pi^-$ as pairs of particles,
the minimum \pt, and the minimum $\chi^2_{\rm IP}$ of each pair are used. The signal sample used for BDT
training is based on simulation, while the background sample uses the sideband $200-250$~MeV above the \Bsb
mass peak from 1/3 of the available data. The BDT is then tested on independent samples from the
same sources. The BDT selection is optimized by taking the signal, $S$, and background, $B$,
events within $\pm$20~MeV of the \Bsb
peak from the preselection and maximizing $S^2/(S + B)$ by using the signal and background efficiencies
provided as a function of BDT. 
 
\begin{figure}[htb]
\centering
\includegraphics[width=0.6\textwidth]{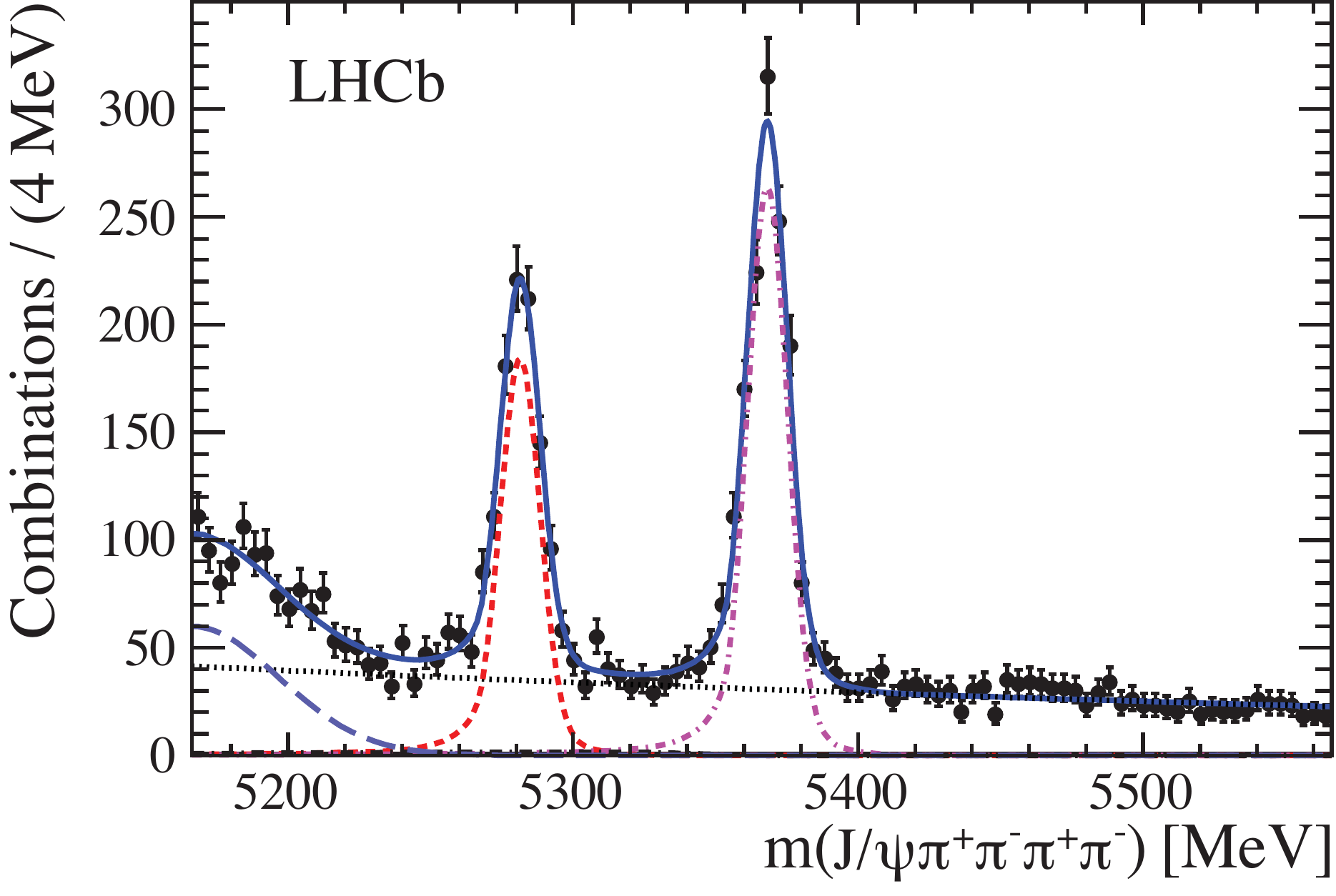}
\caption{\small Invariant mass distribution for $\jpsi \pi^+\pi^-\pi^+\pi^-$ combinations. The data are
fit with Crystal Ball functions for \Bzb [(red) dashed curve] and \Bsb [(purple) dot-dashed curve] signals, an
exponential function for combinatoric background (black) dotted, and a Gaussian shape for
lower mass background (blue) long-dashed. The total is shown with a (blue) solid curve.}
\label{fitdata}
\end{figure}

The $\jpsi\pi^+\pi^-\pi^+\pi^-$ invariant mass distribution is shown in Fig.~\ref{fitdata}. Multiple combinations are at the 6\% level and
a single candidate is chosen based on vertex $\chi^2$ and \jpsi mass. We fit the mass distribution
using the same signal function shape for both \Bsb and \Bzb peaks. This shape is a double
Crystal Ball function \cite{Skwarnicki:1986xj} with common means and radiative tail parameters obtained from
simulation. The combinatorial background is parametrized with an exponential function. There are
1193$\pm$46 \Bsb  and 839$\pm$39 \Bzb decays.
Possible backgrounds caused by particle misidentification, for example $\Bzb\to\jpsi\pi^+K^-\pi^+\pi^-$ decays, would appear as signal if the particle identification incorrectly assigns the $K^-$ as a $\pi^-$. In this case the invariant mass is always below the \Bzb signal region. Evaluating all
such backgrounds shows negligible contributions in the signal regions. These and other low-mass backgrounds are described by a Gaussian distribution.
 
\begin{figure}[b]
\centering
\includegraphics[width=0.99\textwidth]{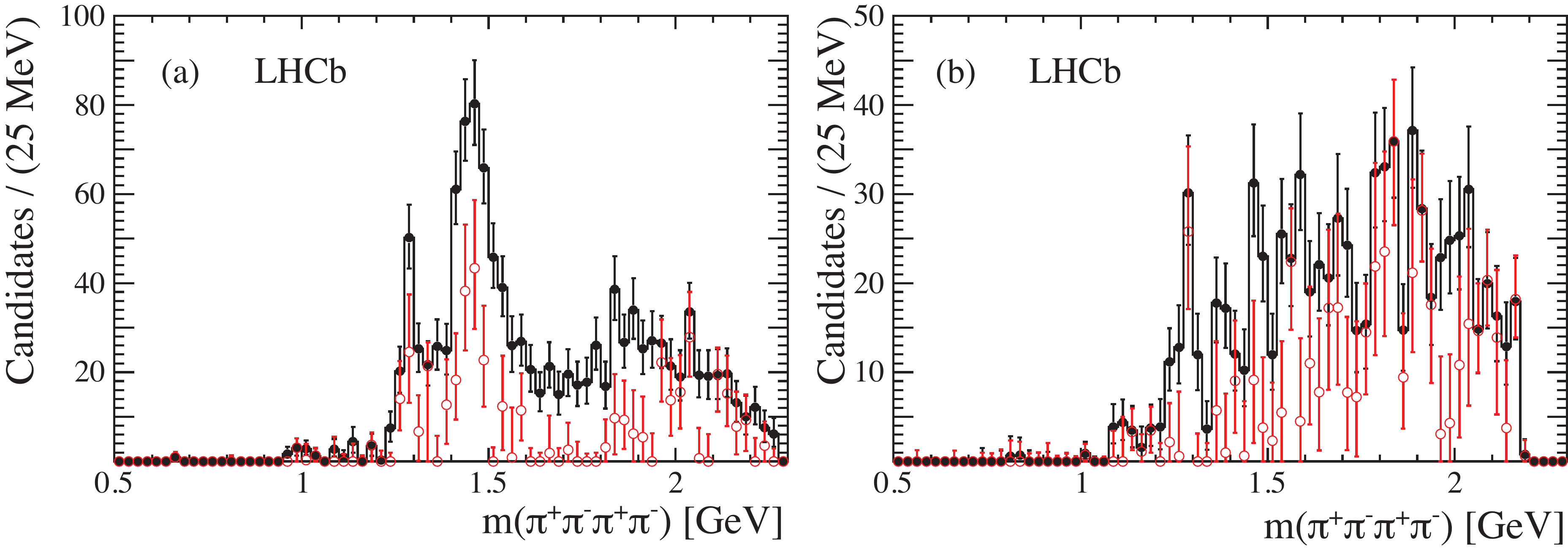}
\caption{\small Background subtracted invariant mass distributions of the four pions in (a)
\Bsb and (b) \Bzb decays are shown in the histogram overlaid with the (black) filled points
with the error bars indicating the uncertainties. The open (red) circles show the helicity $\pm$1 components of the signals.}
\label{m4pi_hel-25}
\end{figure}

In order to improve the four-pion mass resolution we kinematically fit each candidate with the constraints that the $\mu^+\mu^-$ be at the \jpsi mass and that the $\jpsi  \pi^+\pi^-\pi^+\pi^-$ be at the $\Bbar$ mass. The four-pion invariant mass distributions for \Bsb and \Bzb
decays within $\pm$20~MeV of the $\Bbar$ mass peaks are shown in Fig.~\ref{m4pi_hel-25}. The backgrounds, determined from fits to the number of events in the region $40-80$ MeV
above the \Bsb mass, are subtracted. 

There are clear signals around 1285 MeV in both \Bsb and \Bzb decays with structures
at higher masses. The \jpsi
decay angular distribution is used to probe the spin of the recoiling four-pion system. We examine the  distribution of the helicity angle $\theta$ of the $\mu^+$ with respect to the $\Bbar$ direction in the \jpsi rest frame, after correcting for the angular acceptance using simulation. The resulting distribution is then fit by the sum of  shapes  $(1-\alpha)\sin^2\theta$ and $\alpha(1+\cos^2\theta)/2$, where $\alpha$ is the fraction of the helicity $\pm$1 component.  For scalar four-pion states the \jpsi helicity is 0, while for higher
spin states it is a mixture of helicity 0 and helicity $\pm$1 components. We also show in
Fig.~\ref{m4pi_hel-25} the helicity $\pm$1 yields. In the region near 1285 MeV there is a significant helicity $\pm$1
component, as expected if the state we are observing is the \f1 .

There is also a large and wider peak near 1450 MeV in the \Bsb channel. Previously we observed a structure at a mass near 1475 MeV using $\Bsb\to\jpsi  \pi^+\pi^-$ decays that we attributed to $f_0(1370)$ decay. However it could equally well be the
$f_0(1500)$ meson, an interpretation favored by Ochs \cite{Ochs:2013gi}. While the $f_0(1500)$ is known to decay
into four pions, the structure observed in our data cannot be pure spin-0 because of the
significant helicity $\pm$1 component in this mass region. We do not pursue further the
composition of the higher mass regions in either \Bsb or \Bzb decays in this Letter.

We use the measured branching fractions of $\Bsb\to\jpsi  \pi^+\pi^-$ \cite{LHCb:2012ae} and $\Bzb\to\jpsi  \pi^+\pi^-$ \cite{Aaij:2013zpt}  for normalizations. The data selection is updated from that used in previous publications to more closely follow the procedure in this analysis.
 We find signal yields of  22\,476$\pm$177 \Bsb events and 16\,016$\pm$187 \Bzb events within $\pm$20~MeV of the signal peaks. 
The overall efficiencies determined by simulation
are (1.411$\pm$0.015)\% and (1.317$\pm$0.015)\%, respectively, for \Bsb   and \Bzb decays, where the uncertainty is
statistical only. The relative efficiencies for the $\jpsi  \pi^+\pi^-\pi^+\pi^-$ final states with respect to
$\jpsi  \pi^+\pi^-$ are 14.3\% and 14.5\% for \Bsb and \Bzb  decays, with small statistical uncertainties.
We compute the overall branching fraction ratios
\begin{align}
&{\cal{B}}(\Bsb\to\jpsi \pi^+\pi^-\pi^+\pi^-)/{\cal{B}}(\Bsb\to\jpsi \pi^+\pi^-)=0.371\pm 0.015\pm 0.022, \nonumber\\
&{\cal{B}}(\Bzb\to\jpsi \pi^+\pi^-\pi^+\pi^-)/{\cal{B}}(\Bzb\to\jpsi \pi^+\pi^-)= 0.361\pm 0.017\pm 0.021. \nonumber
\end{align}
The systematic uncertainties arise from the decay model (5.0\%), background shape (0.8\%),
signal shape (0.8\%), simulation statistics (1.9\%), and tracking efficiencies (2.0\%), resulting
in a total of 5.8\%.

\begin{figure}[t]
\centering
\includegraphics[width=0.999\textwidth]{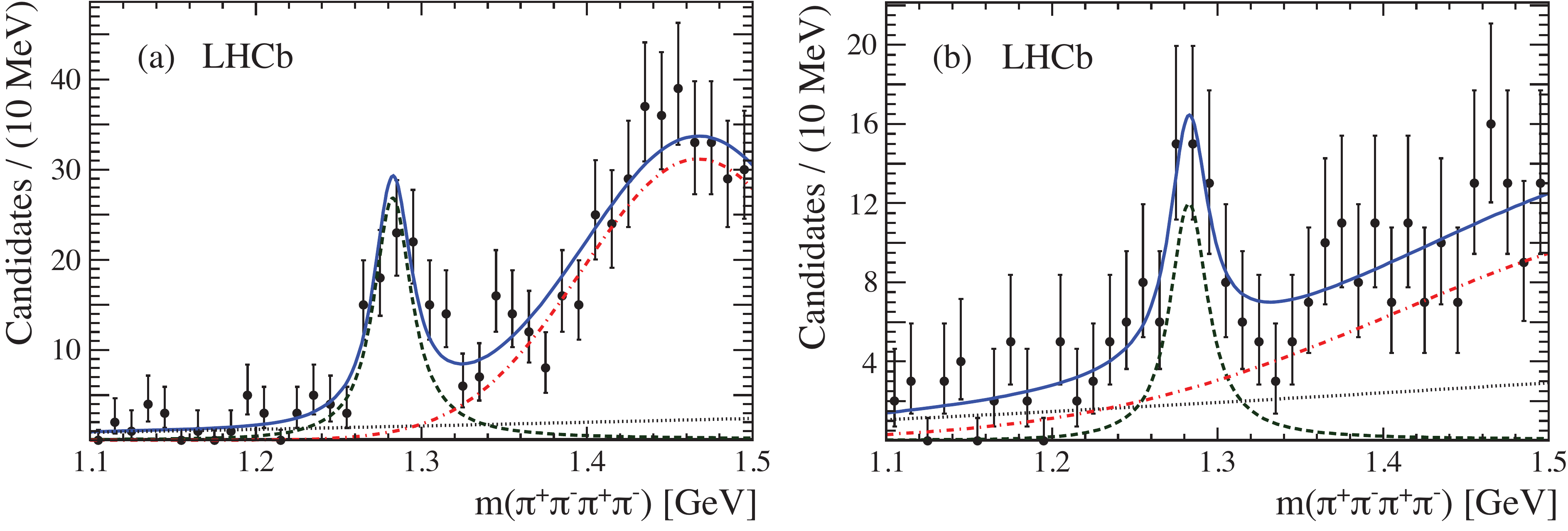}
\caption{\small Fits to the four-pion invariant mass in (a) \Bsb and (b) \Bzb decays. The data
are shown as points, the signals components as (black) dashed curves, the combinatorial
background by (black) dotted curves, and the higher mass resonance tail by (red) dot-dashed curves. }
\label{fitB-f1-fix}
\end{figure}

We proceed to determine the $\jpsi \f1 $ yields by fitting the individual four-pion mass
spectra in both \Bsb and \Bzb final states. The \f1 state is modeled by a relativistic
Breit-Wigner function multiplied by phase space and convoluted with our mass resolution
of 3 MeV. We take the mass and width of the \f1  as 1282.1$\pm$0.6~MeV
and 24.2$\pm$1.1~MeV, respectively \cite{PDG}. The combinatorial background is constrained from
sideband data and is allowed to vary by its statistical uncertainty. Backgrounds from higher
mass resonances are parameterized by Gaussian shapes whose masses and widths are allowed to vary. We restrict the
fits to the interval 1.1$-$1.5~GeV, which contains 94.3\% of the signal. The fits to the data are shown in Fig.~\ref{fitB-f1-fix}. The results of the fits are listed in Table~\ref{tab:f1fit} along with twice the negative change in the logarithm of the likelihood ($-2\Delta\ln  L$) if fit without the signal, and the
resulting signal significance. The systematic uncertainties from the signal shape and higher mass resonances have been included.  Both final states are seen with significance above five standard
deviations. This constitutes the first observation of the \f1 in $b$-hadron decays. As a
consistency check, we also perform a simultaneous fit to both \Bsb and \Bzb samples letting
the mass and width vary in the fit. We find the mass and width of the \f1 to be 1284.2$\pm$2.2 MeV and 32.4$\pm$5.8 MeV, respectively, where the uncertainties are statistical only, consistent with the known values. To determine the systematic uncertainty in the yields we redo the
fits allowing $\pm 1\sigma$ variations of the mass and width values independently. We assign
$\pm$2.7\% and $\pm$2.0\% for the systematic uncertainties on the \Bsb and \Bzb yields, respectively, from this source.

We obtain the branching fraction ratios, using an efficiency of 0.1820$\pm$0.0036\%, determined by simulation, for the $\jpsi \f1 $ final state as
\begin{align}
&\frac{{\cal{B}}(\Bsb\to\jpsi f_1(1285),~f_1(1285)\to\pi^+\pi^-\pi^+\pi^-)}{{\cal{B}}(\Bsb\to\jpsi \pi^+\pi^-)}  = (3.82\pm 0.52^{\,+0.29}_{\,-0.32})\%, \nonumber\\
&\frac{{\cal{B}}(\Bzb\to\jpsi f_1(1285),~f_1(1285)\to\pi^+\pi^-\pi^+\pi^-)}{{\cal{B}}(\Bzb\to\jpsi \pi^+\pi^-)}  = (2.32\pm 0.54\pm 0.11)\%,\nonumber\\
&\frac{{\cal{B}}(\Bdb\to\jpsi f_1(1285))}{{\cal{B}}(\Bsb\to\jpsi f_1(1285))}  =(11.6\pm 3.1^{\,+0.7}_{\,-0.8})\%. \nonumber
\end{align}
\renewcommand{\arraystretch}{1.1}
\begin{table}[t]
\centering
\caption{\small Fit results for $\Bsb\to\jpsi f_1(1285)$ and $\Bzb\to \jpsi f_1(1285)$ decays. }
\vspace{0.2cm}
\begin{tabular}{cccc}\hline
 & Yield & $-2\Delta\ln L$ & Significance ($\sigma$)\\\hline
\Bsb & $110.2\pm15.0$ & 58.1 & 7.2 \\
\Bzb & ~$\,49.2 \pm 11.4$ & 29.5 & 5.2\\
\hline
\end{tabular}
\label{tab:f1fit}
\end{table}
\renewcommand{\arraystretch}{1}

\noindent For the latter ratio we use a $\Bsb/\Bzb$ production ratio of  0.259$\pm$0.015 \cite{fsfd,*Aaij:2013qqa}; this uncertainty is taken as systematic.  The other systematic uncertainties are listed in Table~\ref{tab:f1sys}. The shape of the high-mass tail is changed in the case of \Bsb decays from a single Gaussian to two relativistic Breit-Wigner shapes corresponding to the mass and width values of the $f_1(1420)$ and the $f_0(1500)$ mesons. 
For the \Bzb high mass shape we change from a Gaussian shape to a second order polynomial. The decay model reflects the allowed variation in the fraction of $\rho^0\rho^0$ and $\rho^0\pi^+\pi^-$ decays. The total uncertainties are ascertained by adding the individual components in quadrature separately for the positive and negative values.
\renewcommand{\arraystretch}{1.1}
\begin{table}[t]
\centering
\caption{\small Systematic uncertainties of the branching fractions ${\cal{B}}(\Bbar\to\jpsi \f1 ,~ f_1(1285)\to\pi^+\pi^-\pi^+\pi^-)$ and the $\Bzb/\Bsb$ rate ratio.  The ``+" and ``--" signs indicate the positive and negative uncertainties, respectively. All numbers are in (\%).}
\vspace{0.2cm}
\begin{tabular}{lrrrrrr}\hline
Source& \multicolumn{2}{c}{\Bdb} & \multicolumn{2}{c}{\Bsb}  & \multicolumn{2}{c}{Ratio} \\
      &  + & --  & + &  -- &  + & --\\\hline
Mass \& width of $f_1$ & 2.0 &2.0 & 2.7 &2.7 &1.5 & 1.5\\
Shape of high mass & 0.6 &0&0&3.7&0&3.8\\
Efficiency & 2.0&2.0&2.0&2.0& 0&0\\
Tracking &  2.0&2.0&2.0&2.0& 0&0\\
Simulation statistics & 2.0 & 2.0& 2.0&2.0&0&0\\\hline
Total & 4.0&4.0&4.4&5.7&1.5&4.1\\
\hline
\end{tabular}
\label{tab:f1sys}
\end{table}
\renewcommand{\arraystretch}{1}

Considering the $f_1(1285)$ as a mixed $q\bar{q}$ state, we characterize the mixing with a 2$\times$2 rotation matrix containing  a single parameter, the angle $\phi$, so that the wave functions of the $f_1(1285)$ and its partner, indicated by $f_1^*$, are given by
\begin{eqnarray}
  \label{eq:fmix}
 \ket{f_1(1285)}&=&\cos\phi\ket{n\bar{n}}-\sin\phi\ket{s\bar{s}},\nonumber\\
  \ket{f_1^*}&=&\sin\phi\ket{n\bar{n}}+\cos\phi\ket{s\bar{s}},\nonumber\\
  {\rm where~} \ket{n\bar{n}}&\equiv&\frac{1}{\sqrt{2}}\left(\ket{u\bar{u}}+\ket{d\bar{d}}\right).
\end{eqnarray}
The decay widths can be written as  \cite{Stone:2013eaa}
\begin{eqnarray}
\Gamma(\Bzb\to \jpsi f_1(1285))&=&0.5|A_0|^2|V_{cd}|^2\Phi_0  \cos^2\phi, \nonumber\\
\Gamma(\Bsb\to \jpsi f_1(1285))&=&|A_s|^2|V_{cs}|^2\Phi_s \sin^2\phi,
\end{eqnarray}
where $A_i$ is the tree level amplitude, $V_{cd}$ and $V_{cs}$ are quark mixing matrix elements, and $\Phi_i $ are phase space factors.
The amplitude ratio $|A_0|/|A_s|$ is taken as unity \cite{Stone:2013eaa}.
The width ratio is given by
\begin{equation}
\frac{{\cal{B}}(\Bzb\to \jpsi f_1(1285))}{{\cal{B}}(\Bsb\to \jpsi f_1(1285))}=\frac{\tau_0}{2\tau_s}\frac{|V_{cd}|^2\Phi_0 \cos^2\phi}{|V_{cs}|^2\Phi_s \sin^2\phi},
\end{equation}
where $\tau_s$ is the \Bsb lifetime and $\tau_0$ is the \Bzb lifetime.
The angle $\phi$ is then given by
\begin{equation}
\label{eq:tansqphi}
\tan^2\phi=\frac{1}{2}\frac{{\cal{B}}(\Bsb\to \jpsi f_1(1285))}{{\cal{B}}(\Bzb\to \jpsi f_1(1285))}
\frac{\tau_0}{\tau_s}\frac{|V_{cd}|^2}{|V_{cs}|^2}\frac{\Phi_0}{\Phi_s}=0.1970\pm 0.053^{\,+0.014}_{\,-0.012}.
\end{equation}
The ratio of the phase space factors $\Phi_0/\Phi_s$ equals 0.855.
The other input values are  $\tau_s=1.508$~ps \cite{Aaij:2013oba}, $\tau_0=1.519$~ps, $|V_{cd}|=0.2245$, and $|V_{cs}|=0.97345$ \cite{PDG}. 
We use the lifetime measured in $\Bsb\to\jpsi \phi$ decays as the helicity components are in approximately the same ratio as in $\jpsi f_1(1285)$. No uncertainties are assigned on these quantities as they are much smaller than the other errors.
The resulting mixing angle is
\begin{equation}
\phi=\pm(24.0^{\,+3.1\,+0.6}_{\,-2.6\,-0.8})^{\circ}. \nonumber
\end{equation}
The systematic uncertainty is computed from the systematic errors assigned to the branching fractions.

The $f_1(1285)$ mixing angle has been estimated assuming that it is mixed with the $f_1(1420)$ state.  Yang finds $\phi=\pm(15.8^{\,+4.5}_{\,-4.6})^{\circ}$ using radiative decays \cite{Yang:2010ah}, consistent with an earlier determination of  $\pm(15^{\,+\;\;5}_{\,-10})^{\circ}$  \cite{Gidal:1987bn}.
   A lattice QCD analysis gives  $(31\pm 2)^{\circ}$, while an another phenomenological calculation gives a range between $(20-30)^{\circ}$ \cite{Dudek:2011tt,*Dudek:2013yja,*Close:1997nm}; see also Ref.~\cite{Li:2000dy,*Cheng:2011pb} for other theoretical predictions. In this analysis we do not specify the other mixed partner.
   
If the $f_1(1285)$ is a tetraquark state its wave function would be 
$\ket{f_1}=\frac{1}{\sqrt{2}}\left([su][\bar{s}\bar{u}]+[sd][\bar{s}\bar{d}]\right)$ in order for it to be produced significantly in both \Bsb and \Bzb decays into $\jpsi f_1(1285)$ decays. Using this wave function,
the tetraquark model described in Ref.~\cite{Stone:2013eaa} predicts 
\begin{equation}
\frac{{\cal{B}}(\Bzb\to \jpsi f_1(1285))}{{\cal{B}}(\Bsb\to \jpsi f_1(1285))}=\frac{1}{4}\frac{\tau_0}{\tau_s}\frac{|V_{cd}|^2\Phi_0}{|V_{cs}|^2\Phi_s}=1.14\%,
\end{equation}
with small uncertainties. Our measurement of this ratio of $(11.6\pm 3.1^{\,+0.7}_{\,-0.8})$\% differs by 3.3 standard deviations from the tetraquark 
interpretation including the systematic uncertainty. 

Branching fraction ratios are converted into branching fractions using the previously
measured rates listed in Table~\ref{tab:oldbr}. We correct the \Bsb rates to reflect the updated value of the \Bsb to \Bzb production fraction of 0.259$\pm$0.015 \cite{fsfd,*Aaij:2013qqa}. We determine
\begin{align}
&{\cal{B}}(\Bsb\to\jpsi\pi^+\pi^-\pi^+\pi^-)=(7.62\pm 0.36\pm 0.64 \pm 0.42)\times 10^{-5},\nonumber \\
&{\cal{B}}(\Bzb\to\jpsi\pi^+\pi^-\pi^+\pi^-)=(1.43\pm 0.08\pm 0.09 \pm 0.06)\times 10^{-5}. \nonumber
\end{align}
where the first uncertainty is statistical, the second and third are systematic, being due to 
the relative branching fraction measurements and the errors in the absolute branching fraction normalization, respectively. For the \Bsb decay this normalization error is due to the uncertainty on the production ratio of \Bsb versus \Bzb and is 5.8\%  \cite{Aaij:2013zpt}. For the \Bzb mode the uncertainty is due to the error of 4.1\% on ${{\cal{B}}(\B^-\to\jpsi K^-)}$ \cite{Aaij:2013orb}.

\renewcommand{\arraystretch}{1.1}
\begin{table}[b]
\centering
\caption{\small Branching fractions used for normalization. }
\vspace{0.2cm}
\begin{tabular}{lcc}\hline
Rate & Value& Ref.\\\hline
\rule{0pt}{2.8ex} $\frac{{\cal{B}}(\Bsb\to\jpsi\pi^+\pi^-)}{{\cal{B}}(\Bsb\to\jpsi\phi)}$&$(19.79\pm 0.47\pm 0.52)$\% &\cite{LHCb:2012ae}\\
\rule{0pt}{2.5ex}${\cal{B}}(\Bzb\to\jpsi\pi^+\pi^-)$&$(3.97\pm 0.09\pm0.11\pm 0.16)\times 10^{-5}$&\cite{Aaij:2013zpt}\\
${{\cal{B}}(\Bsb\to\jpsi\phi)}$&$(10.50\pm 0.13\pm0.64\pm 0.82)\times 10^{-4}$&\cite{Aaij:2013orb}\\
 ${{\cal{B}}(\B^-\to\jpsi K^-)}$&$(10.18\pm 0.42)\times 10^{-4}$&         \cite{Aaij:2013orb}\\
\hline
\end{tabular}
\label{tab:oldbr}
\end{table}
\renewcommand{\arraystretch}{1.0}
\vspace* {24mm}
In conclusion, we report the first observations of $\Bzb$ and $\Bsb\to\jpsi f_1(1285)$ decays. These are also the first observations of the $f_1(1285)$  meson in heavy quark decays.
We determine
\begin{align}
&{\cal{B}}(\Bsb\to\jpsi f_1(1285),~f_1(1285)\to\pi^+\pi^-\pi^+\pi^-) = (7.85\pm 1.09^{\,+0.76}_{\,-0.90}\pm 0.46)\times 10^{-6}, \nonumber\\
&{\cal{B}}(\Bzb\to\jpsi f_1(1285),~f_1(1285)\to\pi^+\pi^-\pi^+\pi^-) = (9.21\pm 2.14\pm 0.52\pm 0.38)\times 10^{-7}, \nonumber\\
&{\cal{B}}(\Bsb\to\jpsi f_1(1285))=(7.14\pm 0.99^{\,+0.83}_{\,-0.91}\pm 0.41)\times 10^{-5}, \nonumber\\
&{\cal{B}}(\Bzb\to\jpsi f_1(1285))=(8.37\pm 1.95^{\,+0.71}_{\,-0.66}\pm 0.35)\times 10^{-6}, \nonumber
\end{align}
where we use the known branching fraction ${\cal{B}}(f_1(1285)\to\pi^+\pi^-\pi^+\pi^-)=(11.0^{\,+0.7}_{\,-0.6})$\% \cite{PDG}.
Investigation of  $\Bsb$ and \Bzb decays into $\jpsi\pip\pim\pip\pim$ has revealed the presence of the $\jpsi f_1(1285)$ state in both decay channels. This allows determination of the $f_1(1285)$ mixing angle to be $\pm(24.0^{\,+3.1+0.6}_{\,-2.6\,-0.8})^{\circ}$, even though the mixing companion of this state is not detected. According to Ref.~\cite{Stone:2013eaa}, our measured value disfavors the interpretation of the \f1 as a tetraquark state.
\vskip 4mm

We express our gratitude to our colleagues in the CERN
accelerator departments for the excellent performance of the LHC. We
thank the technical and administrative staff at the LHCb
institutes. We acknowledge support from CERN and from the national
agencies: CAPES, CNPq, FAPERJ and FINEP (Brazil); NSFC (China);
CNRS/IN2P3 and Region Auvergne (France); BMBF, DFG, HGF and MPG
(Germany); SFI (Ireland); INFN (Italy); FOM and NWO (The Netherlands);
SCSR (Poland); MEN/IFA (Romania); MinES, Rosatom, RFBR and NRC
``Kurchatov Institute'' (Russia); MinECo, XuntaGal and GENCAT (Spain);
SNSF and SER (Switzerland); NAS Ukraine (Ukraine); STFC (United
Kingdom); NSF (USA). We also acknowledge the support received from the
ERC under FP7. The Tier1 computing centres are supported by IN2P3
(France), KIT and BMBF (Germany), INFN (Italy), NWO and SURF (The
Netherlands), PIC (Spain), GridPP (United Kingdom). We are thankful
for the computing resources put at our disposal by Yandex LLC
(Russia), as well as to the communities behind the multiple open
source software packages that we depend on.

\newpage
\addcontentsline{toc}{section}{References}
\setboolean{inbibliography}{true}
\bibliographystyle{LHCb}
\bibliography{ANA/jpsi-4pi}

\end{document}